**General Resolution Enhancement Method in Atomic Force Microscopy (AFM) Using Deep Learning**

*Yue Liu, Qiaomei Sun, Wanheng Lu, Hongli Wang, Yao Sun, Zhongting Wang, Xin Lu, Kaiyang Zeng\**

Y. Liu, Q. M. Sun, Dr. W. H. Lu, Dr. H. L. Wang, Y. Sun, Z. T. Wang, X. Lu, Prof. K. Y. Zeng

Department of Mechanical Engineering,
National University of Singapore,
9 Engineering Drive 1, 117576, Singapore
E-mail: mpezk@nus.edu.sg

Keywords: nanoscale topography, atomic force microscopy, deep learning, super resolution, multi-materials network

This paper develops a resolution enhancement method for post-processing the images from Atomic Force Microscopy (AFM). This method is based on deep learning neural networks in the AFM topography measurements. In this study, a very deep convolution neural network is developed to derive the high-resolution topography image from the low-resolution topography image. The AFM measured images from various materials are tested in this study. The derived high-resolution AFM images are comparable with the experimental measured high-resolution images measured at the same locations. The results suggest that this method can be developed as a general post-processing method for AFM image analysis.

**Introduction**

Atomic Force Microscopy (AFM) is a well-known powerful technique to image the surface structures and properties at nanoscale[1,2] with ultrahigh resolution. It tracks cantilever motion affected by the interaction between the tip and the sample surface, therefore, the resolution can reach atomic or molecular level.[3–5] However, the unavoidable experimental errors, such as 'tip crash',[6] the cross-talk between topographic and electrostatic information,[7] the large height variation of the sample surface[8] and the influence by the properties of the sample or the ambient environment[9] can severely reduce the spatial resolution of the AFM images. In addition, scanning a large area with high resolution usually needs long time and may cause the drift of the image and tip wearing as well the distortion of the nanostructures on the sample surface.[10] Generally-speaking, low resolution images certainly contain insufficient information, which may cause some of the important features, including grain boundary, surface defect and dislocation or interface unclear or even ignored. Hence, several methods and techniques were proposed to enhance the resolution and quality of the AFM images, such as by improving the shape and properties of the tip or cantilever,[11–14] the development and application of the multiple frequency excitation techniques,[15–17] the contour metrology[18] and the combination with other microscopy techniques.[19] Generally-speaking, to achieve higher resolution image is a long-term goal in all types of microscopic techniques.

In recent years, owing to the dramatic increase in the computational capabilities and performance as well as the artificial intelligence (AI) technique, machine learning and deep learning (DL) have been implemented in a variety of research fields.[20–22] Especially for image processing, deep learning preforms a strong, effective and efficient capability to detect and classify the objects or features in the images, as well as to predict unknown information based on limited information.[23] For microscopy-based techniques, several applications and methods by using deep learning have been implemented. For example, a DL model developed for Scanning Transmission Electron Microscopy (STEM) can automatically detect and classify the defect transformation.[20] A super-resolution model using DL can restore 3D morphologies of Scanning Electron Microscopy equipped with a Focused Ion Beam (FIB-SEM).[24] For Synchrotron-based X-ray tomography, DL model is also developed for increasing the resolution of the X-ray signals.[25] However, there is limited information on the research of the resolution enhancement using DL methods in the field of AFM image analysis.

Single image super-resolution (SISR) is a multi-solution problem because it uses less information in the low-resolution (LR) images to derive high-resolution (HR) image.[26] Therefore, it is an example-based method which needs a large amount of data (big data) as training samples to learn how to describe the relationships between the LR and HR pixels. A plenty of SISR models using DL have been developed in the recent years.[27–30] Very-Deep Super-Resolution (VDSR) is a mature and popular SISR method with high accuracy and computing speed.[28] It contains very deep convolutional networks which are designed to learn a mapping by end-to-end manner. Furthermore, VDSR requires little pre-processing of the images and engineered features. Instead of preparing LR-HR pairs, VDSR only needs input HR resolution images, which makes it easy to be implemented into microscopical application. Owing to the outstanding performance of DL methods which have been proved on natural photography, videos and some types of microscopy, we propose a new post-processing method to enhance the resolution of the topography images in AFM with a DL model based on VDSR method. In this work, we prepared two training datasets for this model. The first dataset contains topography images measured on only one material. The topography images in the second dataset are obtained from multiple materials. We named them mono-material trained network and multi-material trained network, respectively. Our method requires less limitation on the raw data and can be easy to implement in a wide and general application. In the following sections, the details of the resolution enhancement method will be introduced, and the comparison between the two networks will be illustrated and discussed.

**Results and discussion**

For obtaining a mapping from LR to HR, we employ a very deep learning network constituted of 20 composite layers.[28,31] These 20 layers can be divided into three portions. The first portion is the input layer. The second portion includes 18 middle layers, each of which has a pair of convolutional layer and rectified linear unit (ReLU).[31] The final portion is the reconstruction layer which is constituted of one convolutional layer and a regression layer. The rectifier is an activation function which is considered with principles of biology, it is hence widely used in the deep learning neutral network.[32] ReLU is defined as $f(x) = \max(0, x)$ to filter the features after the convolutional layer. The schematic diagram of this method is shown in **Figure 1**. In this study, we collect a large amount of topography images obtained from AFM as the training dataset and then use it to train the 20 layers convolutional neural network. After training a nonlinear mapping which can detect main features in the images and refine these features, a HR image can be acquired from a LR input image. In addition, this model can be trained in several scale factors simultaneously, which saves the computing resources and time. In this study, three scale factors, 2, 4 and 8, are employed. The initial learning rate is 0.1 which is recommended by the developer of the VDSR model.[28] Both the mono-material network and the multi-materials network are run for 200 epochs.

Due to the fact that different types of materials have dissimilar surface features at nanoscale, establishing dataset for every type of material and training a specific DL network will consume a lot of time and resources. A general DL model that can be applied to the majority of materials is more effective and efficient than a specific model and also beneficial to the future application in the AFM image analysis. Thus, in this study, two training datasets, mono-material dataset and multi-materials dataset, have been prepared to investigate the cross-impact among the multiple types of materials and the viability of the general dataset. The multi-materials dataset contains the same images as in the mono-material dataset and also the collection of the numerous images from other materials. The number of the images from every type of material is not the same. The details of the topography images in the two datasets can be seen in the "Experiments and Method" section.

The two datasets are respectively used to train the same DL model, then two trained networks can be obtained. They are named mono-material network and multi-materials network, respectively. A HR resolution topography image and its simulated LR resolution image obtained by interpolation are employed to evaluate the mono-material network and multi-materials network. The HR resolution images from different methods are shown in **Figure 2(a)**. The resolution has been reduced from 256×256 to 128×128, 64×64 and 32×32, respectively. The left upper corner is extract and enlarged to exhibit the differences of details in the images. Obviously, the HR images obtained from multi-materials and mono-material networks show clearer contour at the grain boundaries. Several features which cannot be recognized in the LR images and bicubic interpolated HR images can be easily found in the DL predicted HR images. In other word, compared with the traditional interpolation method, DL resolution enhancement method performs better on refining tiny features. Meanwhile, there are no significant differences between the HR images predicted by mono-material network and multi-materials network. In order to evaluate the resolution enhancement methods performance quantitatively, two significant parameters, peak-signal-to-noise ratio (PSNR) and structural similarity (SSIM), are employed. A high value of PSNR means a high quality of the reconstructed image,[33] and a high value of SSIM means a high similarity between the reconstructed and the reference images.[34] In **Figures 2(b)** and **(c)**, the quantitative comparisons using PSNR and SSIM illustrate, in most cases, the multi-materials network show a better capability for denoising and restoring the lost information in HR images, whereas the bicubic interpolation shows an inferior capability especially for a large scale factor. For the purpose of investigating the capability of resolution enhancement of all types materials which are involved in the multi-materials network, a test dataset with topography images scanned from multiple materials is used to examine the trained network. The results can be seen in **Table 1**, which demonstrate that the multi-material network has an overall superior performance of resolution enhancement.

From the results mentioned above, it is noted that multi-materials network can adequately restore the details which have been discarded during the artificial resolution decreasing. In order to investigate the capability of resolution enhancement of our DL models in the actual experiment, we repeatedly measure an area of a sample (LAGP) in different resolution. In **Figure 3**, the input image is a raw topography image with 32×32 pixels. It has been enlarged by the scale factors of 2, 4 and 8, respectively and compared with the HR topography images (64×64, 128×128 and 256×256 pixels) on the same area measured by AFM. Owing to the AFM measurement is extremely sensitive and the linear or nonlinear drift usually occurs, there are some unavoidable differences of the topographical features in the repeated experiments and the values of PSNR and SSIM are lower than the simulating test (**Figure 2**), but noted that the HR images predicted by the multi-materials network can clearly show some features which are not shown in the LR image and the bicubic interpolated images. For the scale factors of 4 and 8, the value of PSNR and SSIM between the predicted HR images and their corresponding reference images are higher than those in the bicubic method. The mono-material network performs almost the same as that from the multi-materials network and the quantitative evaluation can be seen in the **Table S1** in the Supporting Information (SI). The quantitative evaluation of the other LR-HR (input-reference) pairs is also shown in the **Table S1 (SI)**. The comparison between the HR topography images and their interpolated LR input images can be seen in the **Table S2 (SI)**, which shows that the simulating resolution enhancement results of these topography images accord with previous test examples (**Figure 2**). The cross-sections analysis is shown in **Figure 4**, which demonstrates that the multi-materials network can smooth the roughness in the LR input image and clearly distinguish the small features of the grains. When the LR image is enlarged by a large scale factor of 8, it still shows an excellent similarity with the HR topography by the AFM measurement. In order to clearly observe the small features restored by the multi-network, we plot 3D surface images of topography which are shown in **Figure S1 (SI)**. The results show the restored HR topography images are similar to the experimental obtained topography images.

Furthermore, we attempt to enhance the resolution of an AFM topography image obtained from a sample of mouse bone. The original image and the predicted HR image can be seen in **Figure S2 (SI)**. By using the multi-material network, the resolution is increased by a scale factor of 4 and some ambiguous features in the LR image can be observed clearly. In fact, it is hard to get an ultrahigh resolution topography image directly by AFM in such type of material because biomaterials generally have extremely complicated structures and these surface structures are easy to be affected or changed by the probing tip during high-resolution scanning.

**Conclusion**

This work introduces a resolution enhancement method based on deep learning neural networks in the AFM topography measurement. A very deep convolutional neural network is employed to derive the high-resolution topography image from the low-resolution topography image, and the derived images are comparable with the experimentally-obtained ones. Due to the fact that the deep learning model is an example-based method, the quality of the training dataset can affect the quality of the network. Therefore, two types of training datasets, mono-material and multi-materials datasets, are involved. The results and analysis developed in this work suggest that both the mono-material network and the multi-materials network can achieve a higher level of accuracy on restoring the subtle features of the sample surface. This method is then verified with refining the compressed low-resolution topography images, it is found that the resulted image highly corresponds to the high-resolution experimental measurement. It is noted that multi-materials dataset does not show any obvious decrease in the resolution which may be related to the cross-impact among a variety of materials, even performs better in some cases. The finding also indicates that to enhance the resolution by establishing a huge multi-materials database including a large number of materials and a deep learning network based on the database can be a general post-processing method for the AFM measurement.

**Experimental and Method Sections**

*Training dataset:* All the topography images are obtained from tapping mode or contact mode using a commercial SPM system (MFP-3D, Oxford Instruments, CA, USA). The mono-material training dataset contains 101 images scanned on LiAlGePO$_4$ (LAGP), a type of ceramic material in bulk shape. The multi-materials training dataset contains 616 images, which are scanned from different types of materials including LAGP, several bulk materials (Pb(Zn$_{1/3}$Nb$_{2/3}$)O$_3$–9%PbTiO$_3$, abalone shell and mouse bone), nanoparticles (LiFePO$_4$ and LiCoO$_2$), thin film materials (doped ZnO, Li$_{1.2}$Co$_{0.13}$Ni$_{0.13}$Mn$_{0.54}$O$_2$, LiMn$_2$O$_4$, Poly(3-hexylthipohene) and several types of doped Poly(vinylidene fluoride))). It should be emphasized that the two datasets have precisely the same images scanned from LAGP. The size and resolution of the raw images are unlimited. The images in our training dataset have different resolutions, such as $128\times128$, $256\times256$ or $512\times512$ pixels, and in a wide range of scanning sizes from 500 nm$\times$500 nm to 10 μm$\times$10 μm. In order to reduce the influences of the experimental-error-induced maximum/minimum points on image contrast, a hard threshold is used to filter the extreme abnormal value. The 2.5$^{th}$ percentile and 97.5$^{th}$ percentile values are set as the threshold value. The filter is defined as:

$$x = F(p, N) = \begin{cases} f(a_{2.5\%}), & p \in [1, a_{2.5\%}], \\ f(p), & p \in (a_{2.5\%}, a_{97.5\%}), \\ f(a_{97.5\%}), & p \in [a_{97.5\%}, N]. \end{cases} \quad (1)$$

where $a_{2.5\%} = \lfloor 2.5\%N \rfloor$, $a_{2.5\%} = \lfloor 97.5\%N \rfloor$; $f(p)$ is the ordered list of one image dataset of raw height value; $x$ is the filtered height value; $p$ is the ordinal number of every point in the dataset. After filtering, the images were normalized by z-scores algorithms. It is defined as:

$$x_{norm} = \frac{x - \mu}{\sigma} \tag{2}$$

where $\mu$ and $\sigma$ is the mean and the standard deviation of the height values in one image, respectively.

*Test dataset:* The test dataset has 18 images which are scanned from the same specimens. The experimental technique and pre-processing of it are also the same as the training dataset.

*VDSR network:* The topography images obtained from AFM have only one channel which can be extracted from the AFM controlling software (IGRO version 6.2) in the text format with high accuracy. Therefore, we developed the input and output processes of the DL model to adapt text files instead of images. To avoid reading and writing images, the modified network hence can be trained on Central Processing Unit (CPU) or Graphic Processing Unit (GPU) of the computation units. This part was implemented in VDSR code designed in MATLAB software (R2018a) with the Neural Network Toolbox, as well as the image pre-processing method. Both training the network and testing the trained network were performed with the MATLAB software. In order to accelerate the speed of the training network, the training dataset was split into small batches, each of which is called minibatch. The size of minibatch in this study is 64. The input images were randomly cut into sub-images with 41×41 pixels which would be stored in the minibatches.

*Quantitative evaluation.* The peak-signal-to-noise ratio (PSNR) is a significant parameter to assess the predicted HR images from bicubic algorithms and our method.[33,35] PNSR (in dB) is defined as:

$$PSNR = 10 \cdot \log_{10}(\frac{MAX_I^2}{MSE}) \quad (3)$$

and

$$MSE = \frac{1}{n}\sum_{i=0}^{n-1}(Y_i - \hat{Y}_i)^2 \quad (4)$$

where $MAX_I$ is the maximum possible pixel value of the image; $MSE$ is the mean squared error between a pair of images both with $n$ pixels; $Y_i$ and $\hat{Y}_i$ is vectors generated from the original images and from the predicted images, respectively. The structural similarity (SSIM)[34] is another effective parameter to evaluate the similarity between two images, which is defined as:

$$SSIM(x, y) = \frac{(2\mu_x\mu_y + c_1)(2\sigma_{xy} + c_2)}{(\mu_x^2 + \mu_y^2 + c_1)(\sigma_x^2 + \sigma_y^2 + c_2)} \quad (5)$$

where $x$ and $y$ represent two images with the same size ($m \times n$ pixels); $\mu_x$ and $\mu_y$ are the averages of $x$ and $y$, respectively; $\sigma_x^2$ and $\sigma_y^2$ are the variance of $x$ and $y$, respectively; $\sigma_{xy}$ is the covariance of $x$ and $y$; $c_1 = (k_1L)^2$ and $c_2 = (k_2L)^2$ are two variables to stabilize the division with weak denominator with $k_1 = 0.01$, $k_2 = 0.03$ and the dynamic range of the pixel-values, $L = 2^{\#bits\ per\ pixel} - 1$, by default.

**Supporting Information**

Supporting Information is available from author upon requirements.

**Acknowledgement**

The manuscript was written through contributions of all authors. YL performed all of the numerical analysis and wrote this manuscript. Other authors were contributed to the large amount of AFM experiments on various materials and discussions about the results and manuscript. All authors have given approval to the final version of the manuscript.
The authors would like to thank the financial support by Ministry of Education, Singapore, through National University of Singapore (NUS) under the Academic Research Fund (AcRF) of grant number R-265-000-596-112. The authors (YL, QMS, YS, ZTW) also thanks the post-graduate scholarship provide by National University of Singapore.

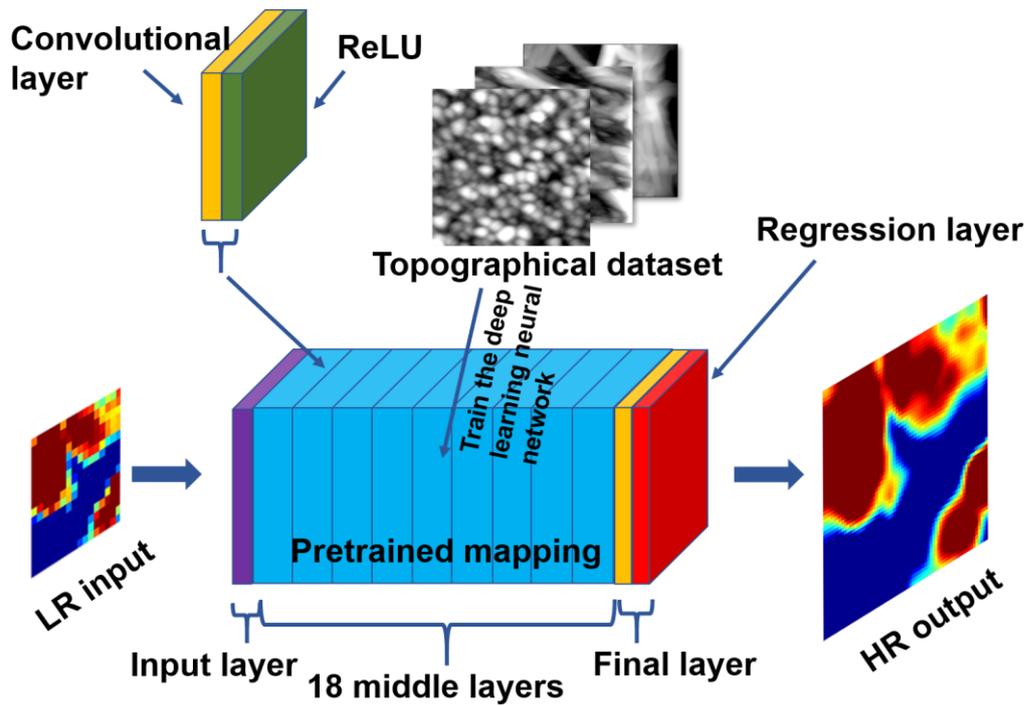

Figure1. Schematic diagram of the resolution enhancement of topography images in AFM using deep learning neural network. The DL convolutional neural network is constituted of four types of single layer, the input layer (purple), the convolutional layer (yellow), ReLU layer (green) and regression layer (red). After training by a reasonable dataset, a HR image can be calculated from a LR input image.

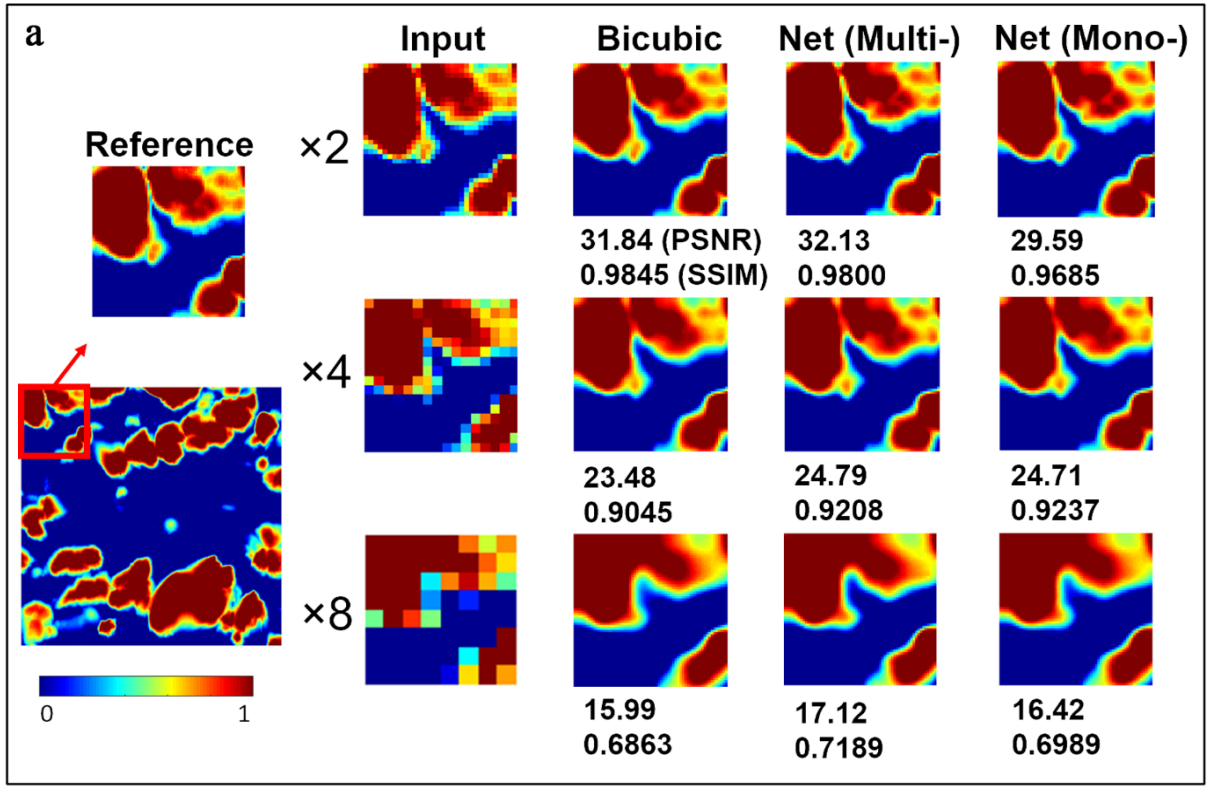
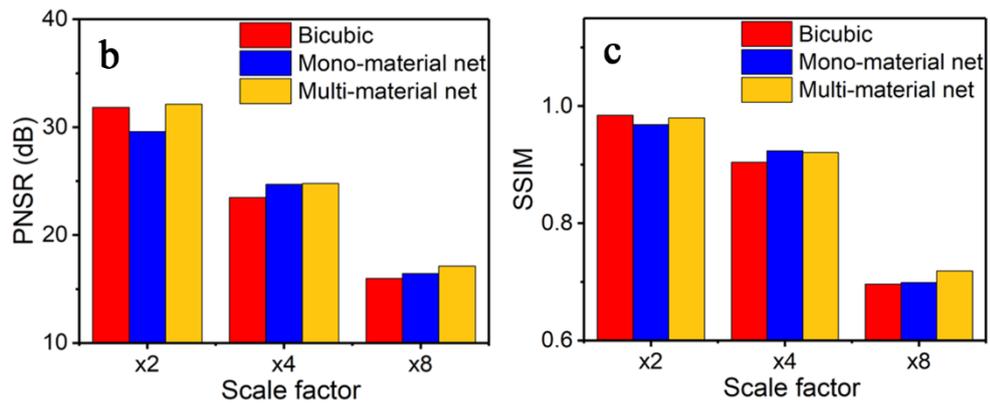

Figure 2. (a) Examples of super-resolution using mono-material and multi-materials networks. The reference topography image is obtained by scanning the sample of LAGP with 256×256 pixels. Its resolution is reduced by scale factors of 2, 4 and 8, respectively, to get the LR input images. The HR images are predicted by bicubic algorithms and our pretrained network, respectively. The performance (PSNR and SSIM) for each predicted image is presented below. (b) Quantitative comparation among bicubic algorithms, mono-material net and multi-material net by PNSR, and (c) by SSIM.

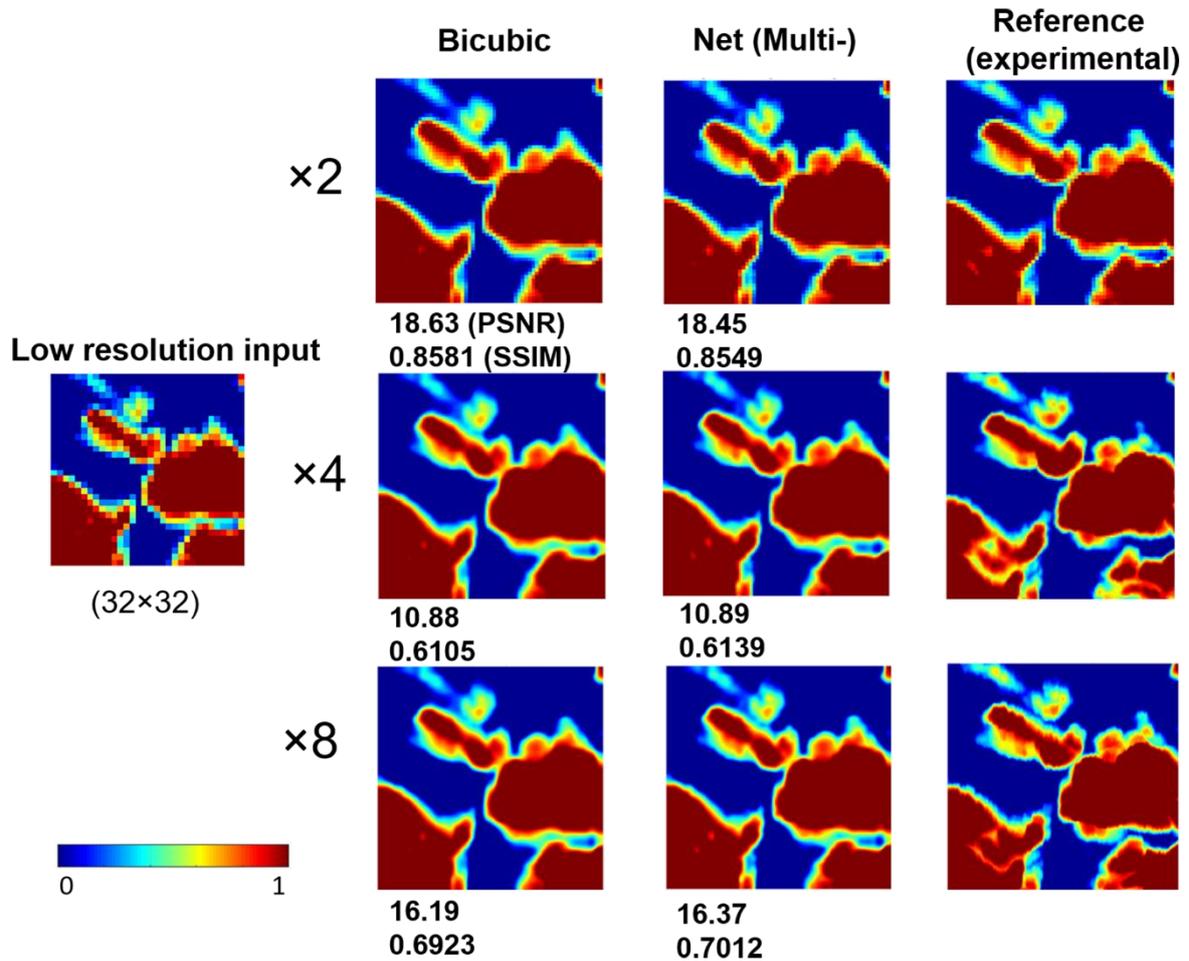

Figure 3. Experimental comparison between the LR scanning and HR scanning on LAGP. The LR input image with 32×32 pixels and HR reference images with 64×64 pixels, 128×128 pixels and 256×256 pixels are all obtained from the AFM measurement. The reconstructed HR images are obtained by using bicubic interpolation and multi-materials network. PSNR and SSIM between the reconstructed HR image and the corresponding reference are shown under each reconstructed HR image.

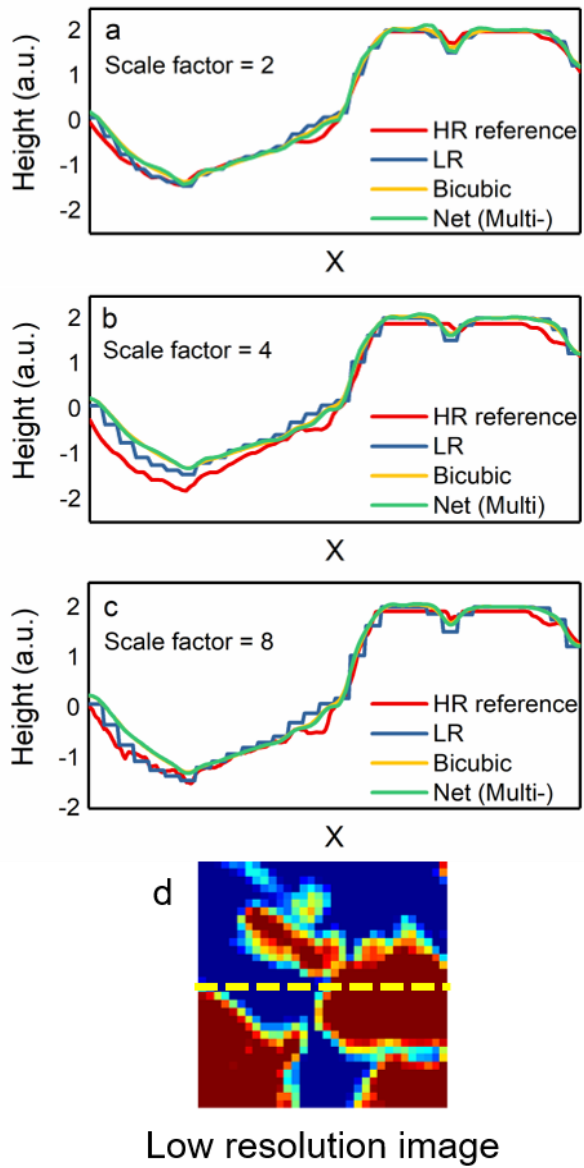

Figure 4. Comparison between the cross-sections of LR image (blue line), HR reference (red line), HR image using bicubic interpolation (yellow line) and HR image using multi-materials network (green line) with different scale factors: (a) 2, (b) 4, and (c) 8, and (d) the spatial position of the picked cross-section is marked by yellow dashed line in the LR image.

Table 1. Quantitative evaluations of the results under different scale factors, represented by the values of PSNR and SSIM, on the test dataset, including the average value and standard deviation of the PSNR and SSIM.

| Scale Factor | | 2<br>[Avg (Std)] | 4<br>[Avg (Std)] | 8<br>[Avg (Std)] |
|---|---|---|---|---|
| Bicubic | PSNR | 27.68(3.54) | 19.66(2.85) | 12.76(2.78) |
|  | SSIM | 0.9697(0.0170) | 0.8226(0.0594) | 0.5018(0.1080) |
| VDSR | PSNR | 28.31(3.20) | 20.90(2.93) | 13.51(2.82) |
|  | SSIM | 0.9692(0.0126) | 0.8584(0.0490) | 0.5400(0.1071) |